Non-perturbative experiments on plasma-mediated particle interaction and the ion wake potential

Ke Qiao,* Zhiyue Ding, Jie Kong, Mudi Chen, Lorin S. Matthews and Truell W. Hyde†
Center for Astrophysics, Space Physics and Engineering Research, Baylor University, Waco, Texas 76798-7310, USA

A non-perturbative method is introduced to measure the particle-particle interaction strengths and *in-situ* confinement for a vertically aligned dust particle pair in a complex plasma. The intrinsic thermal motion of each particle is tracked, allowing the interaction strengths and confinement in both the vertical and horizontal directions to be determined simultaneously. The method is validated through quantitative agreement with previous measurements of the non-reciprocal interaction strength in the vertical direction, the horizontal ion wake attraction, and the decharging and heating of the lower particle when located in the upper particle's wake. The experiment also verifies both theoretical and numerical predictions for the ion wake potential by investigating the ratios among the interaction strengths in the vertical and horizontal directions, as well as in the up- and down-stream directions at varying rf powers. The upstream potential is shown to be asymmetric with unequal screening lengths in the vertical and horizontal directions, implying a subsonic ion flow at low rf powers. Additionally, it is shown that the ratio between the down and up-stream strengths in the vertical direction remains constant at high power, then increases at low power, in agreement with the theoretically predicted increase in the magnitude of the wake potential as the Mach number decreases. Finally, the measured ratio (≈ 5) between the downstream and upstream strengths in the horizontal direction is shown to agree with simulations conducted employing similar plasma parameters.

I. INTRODUCTION

Measurement of particle-particle interactions lies at the heart of research on physical systems ranging from atoms to macroscopic bodies. Experimental limitations on visualization and the lack of separation control for atomic and molecular systems require indirect measurement of these interactions through reconstruction from macroscopic forces [1], photon or neutron scattering [2], or precision vibrational spectroscopy [3]. On the other hand, interaction between medium sized particles (0.01-$10^3$ μm) such as those found in colloidal suspensions, granular materials, and complex (dusty) plasmas can be measured directly through methods such as atomic force microscopy [4] and optical tweezers [5, 6]. A drawback of these methods is that they are intrusive, altering the particle's natural environment and making them unreliable for measurement of the effective interaction between particles mediated by their ambient medium.

In complex plasmas, negatively charged micrometer-sized particles are immersed in partially ionized plasmas, resulting in strong mutual interaction and interesting dynamical behavior such as crystallization [7-9] and ultra-low frequency waves [10-13]. Due to the dynamic response of the ambient plasma to local electric potentials, the medium-mediated interaction plays a particularly important role. For typical complex plasma experiments in a rf discharge, the particles are levitated in the plasma sheath above the lower electrode. The effective interaction between the particles is generally recognized as Debye-Hückel (Yukawa) in the horizontal direction, but is greatly modified in the vertical direction by the ion flow existing in the plasma sheath. The ion flow modified potential (ion wake model) has been the subject of extensive theoretical [14-18] and numerical research [19-25], which has predicted the existence of the ion focus (wake) downstream

*ke_qiao@baylor.edu
†truell_hyde@baylor.edu

of the particle [e.g. 14-16], the effective Yukawa potential with different screening lengths in the upstream and lateral directions [20, 25], and dependence of the wake properties on ion flow speed [15-17, 20-25]. However, experimental verification of these characteristics of the model has proven to be extremely challenging.

Previous measurements of the effective particle-particle interaction in complex plasmas have been made separately in the vertical and horizontal directions. Measurements of the interaction strengths in the vertical direction were conducted by investigation of the dynamic response of a vertical particle pair to a driving voltage applied to the lower electrode [26, 27]. Measurements of the forces in the horizontal direction between vertically aligned dust particles were conducted by either perturbing the particles with a laser [28, 29], or external injection of a new particle [30].

In this research, a non-perturbative method is introduced to measure the particle-particle interaction strength and *in-situ* confinement for a vertically aligned dust particle pair in a complex plasma. The method can be considered an extension of the one reported in [26], where the particle pairs are driven by an external voltage. With no perturbation of the plasma environment, the current method is based solely on the thermal motion of the particles. This seemingly simple extension allows the measurement to be non-intrusive and allows simultaneous measurements in the vertical and horizontal directions (i.e. the directions parallel and perpendicular to the ion flow). The article is organized as follows. First, the method is introduced in Section II, with its theoretical foundation given in II (a) and experimental illustration given in II (b), yielding results that are in quantitative agreement with results from previous experiments that employed different methods. The method is then applied to measure dust particle pairs at varying rf powers as described in Section III, where the ratio among the interaction strengths in the vertical and horizontal directions, as well as in the up- and down-stream directions, are investigated. Discussion on these results in regards to the ion wake potential is given in Section IV. A summary and outlook for future research are given in Section IV.

## II. THE SCANNING MODE SPECTRA (SMS) METHOD

### A. The theoretical foundation

A vertically aligned dust particle pair levitated in a plasma sheath can be modeled as two linearly coupled oscillators, with the equations of motion given as

$$\ddot{x}_1 = -\omega_1^2 x_1 - D_{21}(x_1 - x_2) \quad (1)$$
$$\ddot{x}_2 = -\omega_2^2 x_2 - D_{12}(x_2 - x_1) \quad (2)$$

where $x_1$ and $x_2$ are the displacements of the top and bottom particles from their equilibrium positions, respectively. The frequencies $\omega_1^2$ and $\omega_2^2$ represent their *in situ* confinement at their equilibrium positions, while $D_{12(21)}$ represents the linearized interaction strength of particle 1(2) acting on particle 2(1), defined as the derivative of the interparticle force $F_{12(21)}$ with respect to the interparticle distance $r$ [26]. Since particle motion is due to thermal fluctuations, displacements are small enough that the forces can be considered linear even when confinement at the equilibrium positions is non-harmonic or the particle charge varies with height. Gas friction is balanced by random thermal kick on the particles, which together generates the random but saturated thermal fluctuation of the particle system allowing mode spectra to be obtained. Since these forces do not affect the value of the mode frequencies or eigenvectors [31], they can be neglected in Eqs. (1) and (2).

Two normal modes can be derived from Eqs. (1) and (2), with eigenfrequencies

$$\omega_\pm^2 = \frac{\omega_1^2 + D_{21} + \omega_2^2 + D_{12} \pm \sqrt{(\omega_1^2 - \omega_2^2)^2 + (D_{21} + D_{12})^2 + 2(\omega_1^2 - \omega_2^2)(D_{21} - D_{12})}}{2} \quad (3)$$

where +/- designates the upper and lower frequency. The ratios of the amplitudes of the oscillation of particle 1 and 2 at these frequencies are designated by $\sigma_+$ and $\sigma_-$.

$$\sigma_\pm = \frac{\left((\omega_1^2 - \omega_2^2) + (D_{21} - D_{12}) \mp \sqrt{(\omega_1^2 - \omega_2^2)^2 + (D_{21} + D_{12})^2 + 2(\omega_1^2 - \omega_2^2)(D_{21} - D_{12})}\right)}{2 D_{21}}. \quad (4)$$

Solving Eqs. (3) and (4) for $\omega_1^2, \omega_2^2, D_{12}$ and $D_{21}$ yield the relationships,

$$D_{21}^2 = \frac{(\omega_+^2 - \omega_-^2)^2}{(\sigma_- - \sigma_+)^2} \quad (5)$$

$$\frac{D_{12}}{D_{21}} = -\sigma_+ \sigma_- \quad (6)$$

$$\omega_1^2 = \frac{D_{21}(\sigma_+ + \sigma_-) + (\omega_+^2 + \omega_-^2)}{2} - D_{21} \quad (7)$$

$$\omega_2^2 = \frac{(\omega_+^2 + \omega_-^2) - D_{21}(\sigma_+ + \sigma_-)}{2} - D_{12}. \quad (8)$$

Therefore, if the values of $\omega_+$, $\omega_-$, $\sigma_+$ and $\sigma_-$ can be determined experimentally from the thermal motion of the particles, the interaction strengths $D_{12}$, $D_{21}$ and *in situ* confinements $\omega_1$ and $\omega_2$ can be derived immediately from Eqs. (5-8). Additionally, since this model can be applied to motion in either the vertical or the horizontal direction, the particle interaction and *in situ* confinements in both directions can be determined simultaneously. (Parameters corresponding to the vertical or horizontal direction are denoted by subscript *z* or *x* from here on.)

### B. The experimental illustration

The experiments were carried out in a Gaseous Electronics Conference (GEC) rf reference cell with two 8 cm-diameter electrodes separated by a distance of 1.9 cm. The lower electrode is powered at 13.56 MHz while the upper ring-shaped electrode and chamber act as ground. A 20 mm × 18 mm × 18 mm (height × length × width) glass box placed on the lower electrode creates the confinement potential needed to establish vertical dust pairs. Experiments were conducted in argon plasma at 5.7 Pa employing rf powers of 1.5-10 W. A vertical chain of 8.89-μm melamine formaldehyde (MF) particles was formed within the glass box with the lower particles dropped by decreasing the rf power until only a two-particle vertical pair was left [32, 33]. A vertically fanned laser sheet illuminated the particles and side view images were recorded for 30 s using a CCD camera at 250 fps. The resulting series of images was then analyzed to obtain each particle's position and velocity due to thermal fluctuation.

In standard normal mode analysis, a time series of particle velocities is projected onto the eigenvectors corresponding to the modes theoretically derived from a presumed form of the particle interaction [34-37]. In this method, we instead scan the entire eigenvector space by varying the ratio σ of the oscillation amplitudes of the top and bottom particles from ∞ to -∞ (see Fig. 1(a)). The phase angle $\alpha$ is defined as $\sigma = \cos(\alpha)/\sin(\alpha)$ (where $\alpha$ varies between 0 and π as σ varies between ∞ and -∞). By projecting particle thermal velocity onto each of these eigenvectors and employing a Fourier transformation, we obtain a power spectrum that we refer to as the scanning mode spectrum (SMS). The maxima in the SMS represents the eigenvectors corresponding to the normal modes. Fig. 1(b) and (c) shows the SMS for a particle pair at a rf power of 5.5 W obtained from the vertical and horizontal velocities respectively. As expected from the theory presented in Section II (a), the SMS in both Fig. 1(b) and (c) exhibit two spectral lines, corresponding to the two normal modes with frequencies $\omega_+$ (higher frequency) and $\omega_-$ (lower frequency). The values of $\alpha$ for which the maxima in the two spectral lines occur identify the eigenvectors corresponding to each mode, represented by the ratios of the amplitudes of particle 1 and 2, $\sigma_+$ and $\sigma_-$.

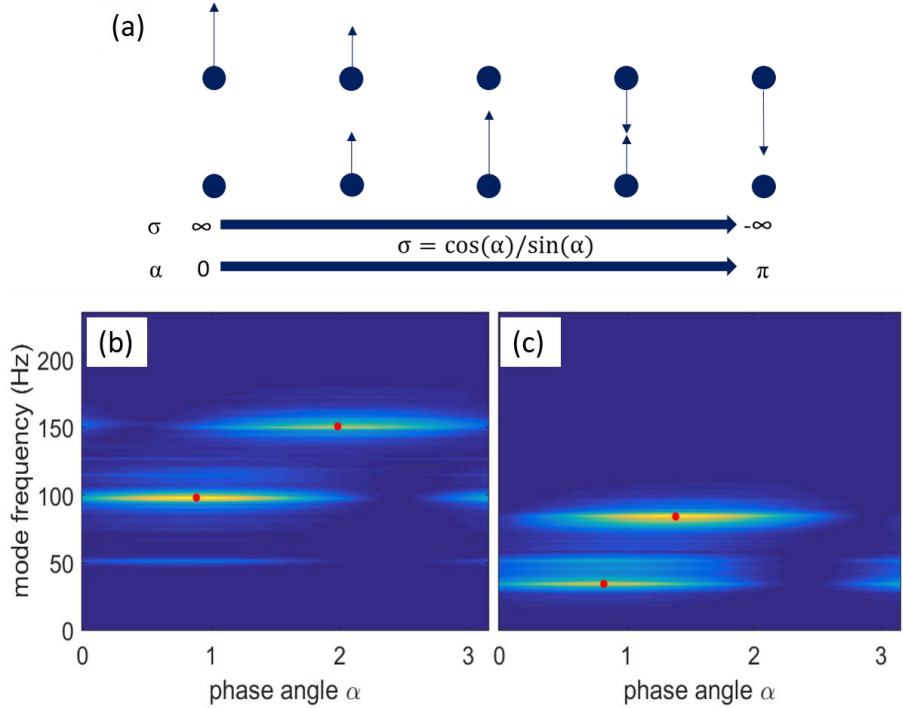

FIG. 1. (Color online) (a) Eigenvector space scanned by varying the ratio σ from ∞ to -∞ (0 ≤ $\alpha$ ≤ $\pi$). (b-c) SMS for a particle pair at rf power of 5.5 W obtained from the (b) vertical and (c) horizontal velocity. Red solid dots indicate the maxima in the spectral lines.

At this point, it is helpful to discuss the SMS in Fig. 1 in comparison to the ideal case of a reciprocal Coulomb potential between the particles. In the ideal case, we have

$$D_{12z} = D_{21z} = \frac{q_1 q_2}{2\pi \varepsilon_0 m r^3} \tag{9}$$

and

$$D_{12x} = D_{21x} = -\frac{q_1 q_2}{4\pi \varepsilon_0 m r^3}, \tag{10}$$

where $q_{1(2)}$ is the particle charge and $m$ is the particle mass. According to Eq. (4), the system will have one sloshing (σ > 0, $\alpha$ < $\pi$/2) and one breathing (σ < 0, $\alpha$ > $\pi$/2) mode in both the $x$ and $z$ directions. If it is assumed further that $\omega_{1z} = \omega_{2z}$ and $\omega_{1x} = \omega_{2x}$, then σ$_{+z}$ = -1, σ$_{-z}$ = 1, σ$_{+x}$ = 1 and σ$_{-x}$ = -1. This means that in the vertical direction, the higher frequency mode is the breathing mode, whereas the higher frequency mode in the horizontal direction is the sloshing mode.

As shown, the SMS obtained from the vertical motion (Fig. 1(b)) exhibits a sloshing mode with $\omega_{-z} = 98$ s$^{-1}$, $\sigma_{-z} = 1.21$ ($\alpha_{-z} = 0.88$) and a breathing mode with $\omega_{+z} = 151$ s$^{-1}$, $\sigma_{+z} = -2.13$ ($\alpha_{+z} = 2.01$). These results are qualitatively similar to the Coulomb case in that the breathing mode is the high frequency mode. Utilizing the relationships provided in Eqs. (5-8), the interaction and *in situ* confinement strengths can be determined, with $D_{12z} = 10124$ s$^{-2}$, $D_{21z} = 3941$ s$^{-2}$, $\omega_{1z}^2 = 10395$ s$^{-2}$ and $\omega_{2z}^2 = 7823$ s$^{-2}$. The resulting interaction strengths $D_{12z}$ and $D_{21z}$ are positive but unequal, with a ratio $D_{12z}/D_{21z} \approx 2.57$. Thus, the interaction between the two particles is mutually repulsive, but nonreciprocal exhibiting a greater strength from the top to lower particle, $D_{12z} > D_{21z}$, in agreement with [26]. An *in-situ* confinement acting on the upper particle greater than that for the lower particle, $\omega_{1z}^2 > \omega_{2z}^2$, was also reported in [26], and explained as the result of the reduction in charge of the bottom particle due to the top particle's ion wake. The same explanation applies here. Since the spatial scale for changes in the electric field at the power setting employed is typically an order of magnitude less than that of the change in charge [38, 39], the ratio between charges $q_2/q_1$ is approximately equal to $\omega_{2z}^2/\omega_{1z}^2$, which is found here to be 0.75, in excellent agreement with the value of 0.78 reported in [26].

The SMS obtained from the horizontal motion (Fig. 1(b)) exhibit two sloshing modes ($\alpha < \pi/2$) with $\omega_{+x} = 83$ s$^{-1}$, $\omega_{-x} = 34$ s$^{-1}$ and $\sigma_{+x} = 4.47$ ($\alpha_{+x} = 1.35$) and $\sigma_{-x} = 1.06$ ($\alpha_{-x} = 0.82$). This finding, which agrees with results reported in [40], is completely different from the Coulomb case, where both a sloshing and a breathing mode are expected. From these measurements, we can obtain the results $D_{12x} = 7984$ s$^{-2}$, $D_{21x} = -1676$ s$^{-2}$, $\omega_{1x}^2 = 1039$ s$^{-2}$ and $\omega_{2x}^2 = 661$ s$^{-2}$. It is important to note that the interaction strength $D_{21x}$ is negative, meaning that the force from the bottom particle acting on the top particle is repulsive (see Eq. (10)). The positive value for $D_{12x}$, however, indicates that the top particle exerts an attractive force on the bottom particle. Since $D_{12x}$ can be considered as the confinement strength on the bottom particle provided by a horizontal potential well positioned directly below the top particle, it serves as a direct verification and quantification of the ion wake effect. Although the ion wake attraction force has been studied previously in both experiment [28-30] and simulation [23, 24], this has only been for cases where the lower particle is displaced from the vertical axis. In this case, we measure the interaction strength (spring constant) $D_{12x}$ directly *on* the z axis. Quantitative comparison with previous research results will be given in the next section.

Another phenomenon of interest is the heating of the lower particle due to the ion flow [40], which may be the origin for more complicated phenomena such as ion flow-induced instabilities in bilayer systems [41, 42]. The SMS method quantitatively represents this heating through values of $|\sigma_{\pm z}|$ and $|\sigma_{\pm x}|$ greater than 1, as evidenced in the current results. Examining Eq. (4) shows that this can be caused by either nonreciprocity in the particle-particle interaction, $D_{12} \neq D_{21}$, or due to a difference in confining strength, $\omega_1 \neq \omega_2$. Since identical confinement acting on particles 1 and 2 ($\omega_1 = \omega_2$) would lead to $\sigma_- = 1$ and $\sigma_+ = -D_{12}/D_{21}$ (Eq. (4)), the slight heating observed in the lower-frequency modes ($\sigma_{-z} = 1.21$, $\sigma_{-x} = 1.06$) serves as direct evidence of the fact that $\omega_1 > \omega_2$, which results from the reduction in charge of the bottom particle as discussed above. It will be shown in the next section that the value of $\sigma_{-z}$ is almost constant for high rf power, eliminating the possibility that the deviation of $\sigma_{-z}$ from 1 is caused by experimental error. Meanwhile the higher-frequency modes exhibit much stronger heating ($\sigma_{+z} = -2.13$, $\sigma_{+x} = 4.47$), primarily caused by the nonreciprocity in particle-particle interaction, since $\sigma_{+z}$ and $\sigma_{+x}$ are close to the values of $-D_{12z}/D_{21z}$ ($\approx -2.57$) and $-D_{12x}/D_{21x}$ ($\approx 4.76$).

### III. RESULTS FROM APPLICATION OF THE SMS METHOD

As shown in the previous section, the results obtained using the SMS method are in quantitative agreement with results obtained employing other experimental methods, verifying the validity of the method. The primary advantage of this method lies in its ability to make measurements in both the vertical (z) and horizontal (x) directions simultaneously. This allows us to apply the SMS measurement under varying rf powers to compare the interaction strengths in the x and z directions ($D_x$ and $D_z$), as well as in the upstream and downstream directions ($D_{21}$ and $D_{12}$). Note that the non-perturbative nature of the measurement only refers to the SMS method itself. As the rf power and particle equilibrium position vary, plasma parameters such as the particle charge $q_{1(2)}$, screening length $\lambda$, confinements $\omega_{1x(z)}^2$ and $\omega_{2x(z)}^2$ and ion flow velocity $u_i$ all change as well. Therefore, our goal in this section is not to map the potential around the dust particles, but rather to extract useful information on the ion wake potential by investigating the ratios among the above-mentioned quantities at specific powers, with the subsequent variation of the plasma parameters taken into account.

As the rf power is decreased from 10 to 1.5 W while keeping the pressure constant at 5.7 Pa, the center of mass of the two particles moves first upward and then downward, reaching a maximum height at rf power $\approx$ 2 W. However, the particle separation $r$, which can be measured directly, continuously increases (Table I). The lower limit for the rf power is 1.5 W, as any further decrease causes the bottom particle to drop to the lower electrode.

As it is difficult, if not impossible, to measure plasma conditions within the box using a Langmuir probe, the particle charge $q$ and screening length $\lambda_x$ are obtained from an independent measurement using the thermally exited mode spectra for a two dimensional (2D) dust cluster at the same experimental conditions [36]. Since the particles are in a horizontal plane, the particles are assumed to interact through a Yukawa potential and their charge and motion are unaffected by the ion wake. Thus the charge which is listed in Table I is assumed to be equal to the charge on the top particle in the vertical pair. As the spectral method is based on the horizontal motion of the particles, the screening length measured is $\lambda_x$ in the horizontal direction. Below a certain critical power, the cluster loses its 2D nature, making these measurements invalid for our assumption of Yukawa-type interaction. For this reason, we place a lower limit on the vertical confinement with the minimum rf power 1.79 W ($\omega_{-z}^2 \approx$ 2000 s$^{-2}$ for vertical confinement, as introduced below).

TABLE I. Experimentally measured interparticle spacing, vertical oscillation frequency, particle charge and horizontal screening lengths for varying rf powers

| rf power (W) | 1.79 | 2.44 | 3.19 | 4.05 | 5.00 | 6.06 | 7.22 | 8.47 | 9.83 |
|---|---|---|---|---|---|---|---|---|---|
| r (µm) ($\pm$5) | 616 | 396 | 315 | 269 | 238 | 220 | 203 | 193 | 181 |
| $\lambda_x$ (µm) ($\pm$50) | 369 | 341 | 301 | 265 | 262 | 252 | 245 | 244 | 240 |
| q (e) ($\pm$500) | 11431 | 11414 | 10451 | 10470 | 10796 | 11351 | 10269 | 9589 | 9372 |
| $\omega_{-z}^2$ (s$^{-2}$) ($\pm$100) | 2029 | 3423 | 5239 | 6859 | 8457 | 10090 | 12009 | 13940 | 15964 |

The SMS method is then applied to measurements on the dust particle pair at varying rf powers. As explained in Section II, $\omega_{+z(x)}^2$, $\omega_{-z(x)}^2$, $\sigma_{+z(x)}$ and $\sigma_{-z(x)}$ are measured directly, and from these the interaction strengths $D_{12z(x)}$, $D_{21z(x)}$ can be derived from Eqs. (5-8). The vertical sloshing mode frequency $\omega_{-z}^2$ can be used to represent the vertical confinement. As can be seen from Eq. (3), $\omega_{-z}^2$ is approximately the average of $\omega_{1z}^2$ and $\omega_{2z}^2$ for small values of $\Delta\omega_z^2$ ($\omega_{1z}^2 - \omega_{2z}^2$). This quantity, which as shown in Table I decreases with decreasing rf power, will be used as the argument for other measured physical quantities being investigated [43] instead of the rf power, as the applied power is related to the details of a particular experimental set-up. The interaction strengths $D_{12z}$, $D_{12x}$, $D_{21z}$ and $-D_{21x}$ ($D_{21x} < 0$) derived for varying confinement strengths $\omega_{-z}^2$ are shown in Fig. 2. These quantities constitute the basis for investigation of the wake-mediated potential between the two grains, as will be seen over the rest of the paper.

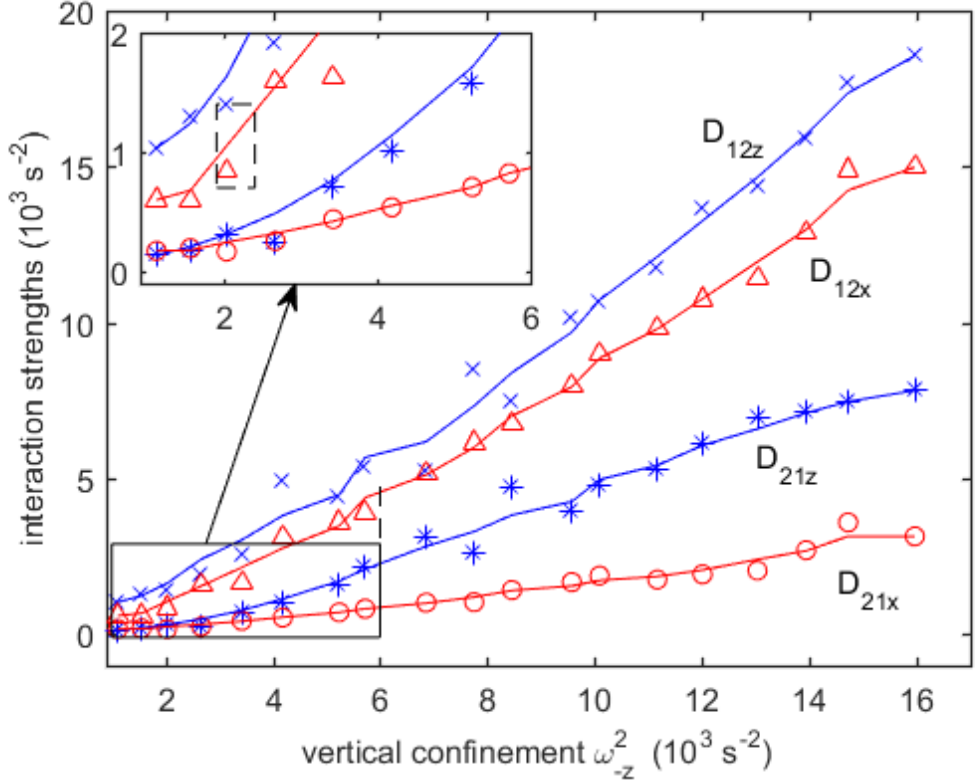

Fig. 2. (Color online) The interaction strengths measured from the SMS method. Inset shows an enlarged view at low confinement strength (i.e. low power). Symbols designate experimentally measured values, while lines show the smoothed data using a moving average filter. The dashed rectangle in the inset marks the area comparable to [30].

A.  Asymmetric Yukawa potential upstream of the particle: ratio of $|D_{21z}/D_{21x}|$

In this sub-section, we will show that a study of the ratio of the interaction strengths in the horizontal and vertical directions, $|D_{21z}/D_{21x}|$ leads to the conclusion that the ion wake causes the shielding length to be asymmetric upstream of a dust grain for weak confinement $\omega_{-z}^2$. The ratio $|D_{21z}/D_{21x}|$, as a function of the vertical confinement strength is shown in Fig. 3. It is calculated by smoothing the data for $D_{21z}$ and $D_{21x}$ (Fig. 2) using a moving average filter before dividing the two values. It can be seen that the ratio $|D_{21z}/D_{21x}|$ fluctuates around a value greater than 2 for $\omega_{-z}^2 > 6000$ s$^{-1}$ and decreases for $\omega_{-z}^2 < 6000$ s$^{-2}$, suggesting a phenomenological division of the range of $\omega_{-z}^2$ into two regions, Region I and II, with a boundary at $\omega_{-z}^2 \approx 6000$ s$^{-1}$. The physical mechanism behind this division will be discussed below.

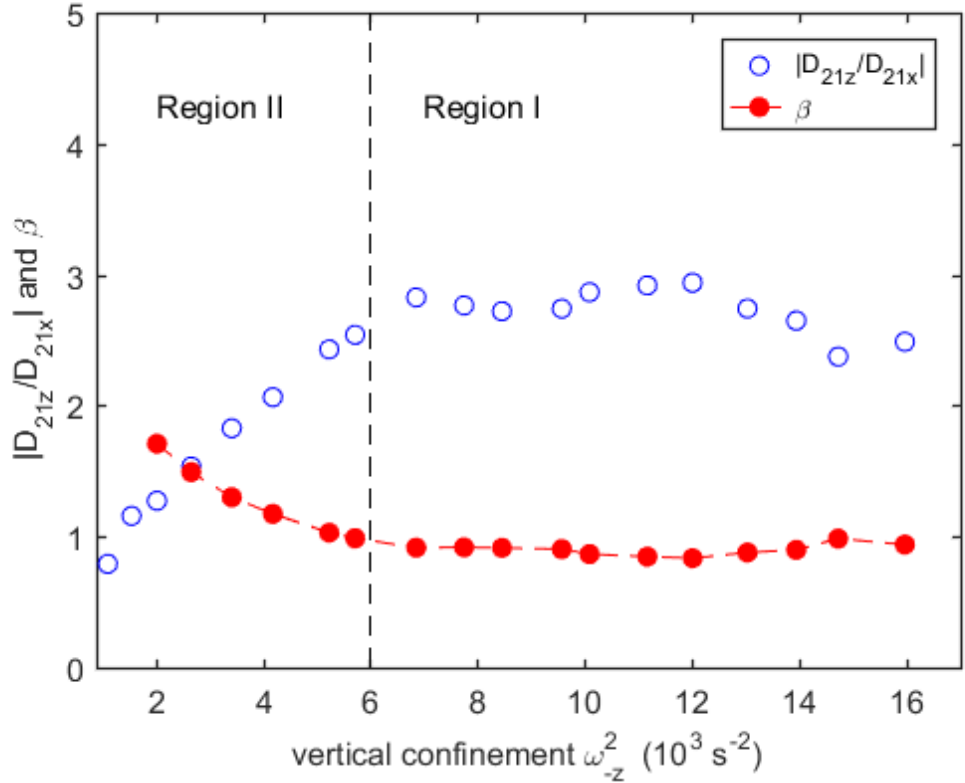

FIG. 3. (Color online) The ratios $|D_{21z}/D_{21x}|$ (calculated from the smoothed data as shown in Fig. 2) and $\beta = \lambda_z/\lambda_x$.

It has been shown theoretically that the upstream potential for a particle in an ion flow can still be modeled effectively by a screened Coulomb (Yukawa) potential [15, 16, 20, 21, 25, 44-46]. A Yukawa potential with equal screening lengths in the upstream and lateral direction ($\lambda_z = \lambda_x$)

$$\phi(r) = \frac{q_1 q_2}{4\pi\epsilon_0 r} \exp(-r/\lambda), \quad (11)$$

will yield the interaction strengths

$$D_{21x} = \frac{1}{m}\frac{\partial f_x}{\partial x} = -\frac{q_1 q_2}{4\pi\epsilon_0} e^{-r/\lambda} \left(\frac{1}{r^3} + \frac{1}{\lambda r^2}\right), \quad (12)$$

and

$$D_{21z} = \frac{1}{m}\frac{\partial f_z}{\partial z} = \frac{q_1 q_2}{4\pi\epsilon_0} e^{-r/\lambda} \left(\frac{2}{r^3} + \frac{2}{\lambda r^2} + \frac{1}{\lambda^2 r}\right). \quad (13)$$

As can be seen by using Eqs (12-13), the ratio $|D_{21z}/D_{21x}|$ is greater than 2, which agrees with the experimentally measured values for $|D_{21z}/D_{21x}|$ in Region I. This is not true for the trend seen for this ratio in Region II. The fact that $|D_{21z}/D_{21x}| < 2$ in Region II can be explained by assuming different screening lengths $\lambda_z$ and $\lambda_x$, for which the interparticle potential is given by

$$\phi(x,z) = \frac{q_1 q_2}{4\pi\epsilon_0 r} \exp\left(-\sqrt{\frac{x^2}{\lambda_x^2} + \frac{z^2}{\lambda_z^2}}\right), \quad (14)$$

This leads to the interaction strengths

$$D_{21x} = \frac{\partial f_x}{m\partial x} = -\frac{q_1 q_2}{4\pi\epsilon_0} e^{-r/\lambda_z} \left(\frac{1}{r^3} + \frac{1}{\lambda_z r^2}\beta^2\right) \quad (15)$$

and

$$D_{21z} = \frac{\partial f_z}{m\partial z} = \frac{q_1 q_2}{4\pi\epsilon_0} e^{-r/\lambda_z}\left(\frac{2}{r^3} + \frac{2}{\lambda_z r^2} + \frac{1}{\lambda_z^2 r}\right), \quad (16)$$

where $\beta \equiv \lambda_z/\lambda_x$. Experimentally measured values of $D_{21z}/D_{21x}$, $\lambda_x$, and the particle separation $r$ (Table I) allow $\beta$ and $\lambda_z$ to be calculated over the range of $\omega_{-z}^2$ using Eqs. (15-16). As shown in Fig. 3, $\beta$ maintains a nearly constant value ≈ 1 in Region I, but increases as the vertical confinement decreases in Region II. The values determined for $\lambda_z$, $\lambda_x$ and the measured interparticle spacing $r$ as a function of vertical confinement strength are shown in Fig. 4. The difference between the values of $\lambda_z$ and $\lambda_x$ in Region II can be clearly seen. Fig. 4 also shows that the increase in $\lambda_z$ is commensurate with that of $r$ in Region II, the physical meaning of which will be discussed below as we investigate the relation between the vertical component of the upstream and downstream interactions, $D_{12z}$ and $D_{21z}$.

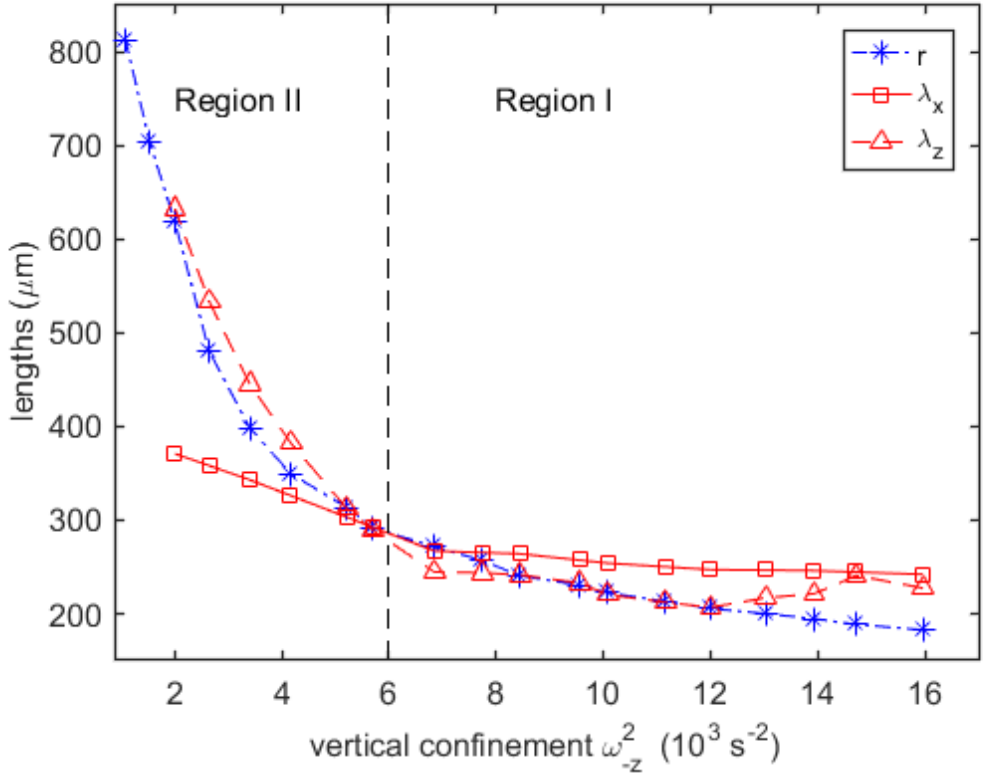

Fig. 4. (Color online) The interparticle spacing $r$ and upstream screening lengths $\lambda_x$ and $\lambda_z$, as functions of the vertical confinement $\omega_{-z}^2$.

### B. Ion wake potential in the vertical direction: ratio of $|D_{12z}/D_{21z}|$

Next, we examine the ion wake potential in the vertical direction by investigating the relationship between $D_{12z}$ and $D_{21z}$. According to linear response theory [e.g. 15, 16], the potential $\phi$ around a particle located in the ion flow consists of two components, i.e. $\phi = \phi_D + \phi_W$, where $\phi_D$ is the screened Coulomb potential which is symmetric above (upstream) and below (downstream) the particle, and $\phi_W$ the (ion) wake potential, which only appears downstream. With the cold ion approximation, these two types of potential are given as

$$\phi_W = \frac{q^2}{\lambda_e} f_W\left(\frac{r}{\lambda_e}, M\right) \tag{17}$$

and

$$\phi_D = \frac{q^2}{\lambda_e} f_D\left(\frac{r}{\lambda_e}, M\right). \tag{18}$$

Both $f_w$ and $f_D$ are integrals in momentum space and depend only on the Mach number $M$ and the normalized interparticle separation $r/\lambda_e$. The integral $f_w$ yields a wake potential $\phi_w$ having an oscillatory structure exhibiting a series of peaks (valleys) while the integral $f_D$ provides $\phi_D$ as a potential exhibiting a classical Yukawa nature [15, 16].

Accordingly, the force $F = F_D + F_W$ and interaction strength (force derivative) $D = D_D + D_W$ have the same symmetry as $\phi$, i.e. the interaction between the two particles $D_D$ is reciprocal, and the wake interaction $D_W \approx 0$ only for the top particle acting on the bottom one. Thus, along the z axis, the screened Coulomb strength $D_D$ can be considered equal in magnitude to the upstream strength $D_{21z}$, and the wake strength $D_W$ equals the difference $D_{12z} - D_{21z}$. In Fig. 2, the values $D_{12z}$ and $D_{21z}$ show a nonreciprocal relationship, $D_{12z} > D_{21z}$ over the entire range of confinement strengths. Therefore the wake strength $D_W = D_{12z} - D_{21z}$ is always positive, corresponding to a repulsive interaction. The ratios of the vertical oscillation amplitudes $\sigma_{-z}$ and $\sigma_{+z}$, shown in Fig. 5, are almost constant across Region I with values $\approx 1.2$ and $2.0$, respectively. (As mentioned in Section II (b), a constant value of $\sigma_{-z} \approx 1.2$ eliminates the possibility that the measured result $\omega_{1z} > \omega_{2z}$ is caused by experimental error.) As a result, the nonreciprocity ratio $|D_{12z}/D_{21z}|$ given by Eq. (6) is also constant $\approx 2.4$ and hence $D_W/D_D \sim 1.4$. In Region II, both $|\sigma_{-z}|$ and $|\sigma_{+z}|$ increase as the confinement decreases. Therefore both $|D_{12z}/D_{21z}|$ (Fig. 5) and $D_W/D_D$ increase. The reason for this, and its interpretation in regards to the wake potential in the vertical direction, will be discussed in Section IV.

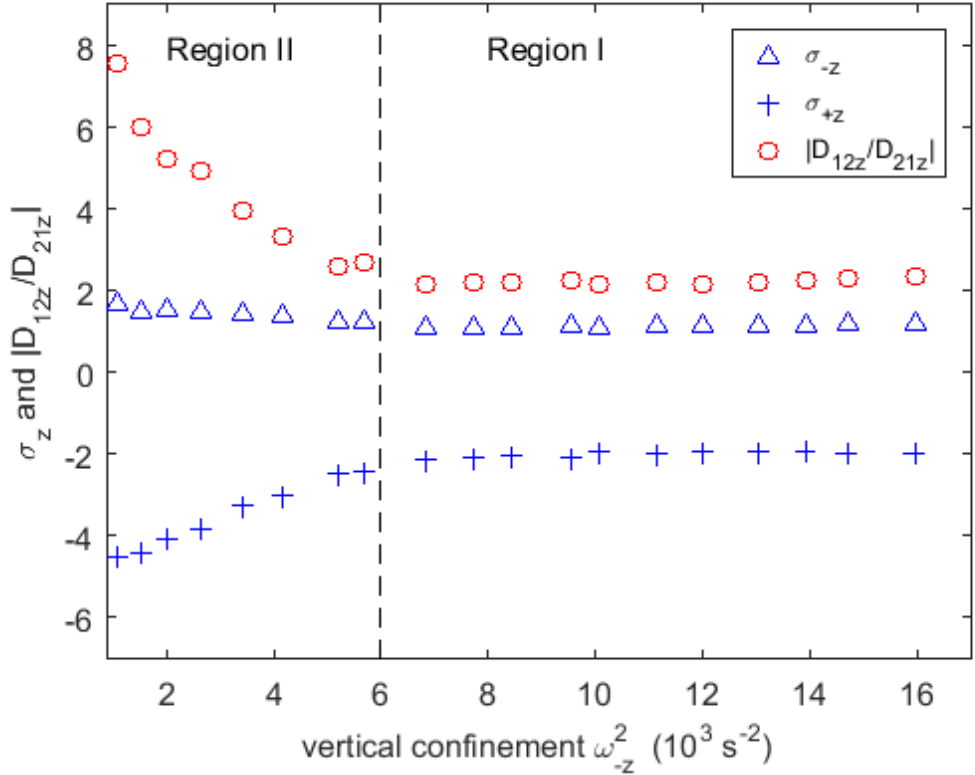

FIG. 5. (Color online) The ratios $\sigma_{+z}$ and $\sigma_{-z}$, corresponding to the two vertical modes, and the ratio of nonrecipocity $|D_{12z}/D_{21z}|$.

C. Ion wake potential in the horizontal direction: ratio of $D_{12x}/D_{21x}$

Finally, we examine the strength of the horizontal wake attraction $D_{12x}$ and compare the down- and up-stream interaction strengths $D_{12x}$ and $D_{21x}$. As shown in Fig. 2, the strength of the horizontal wake attraction $D_{12x}$ varies from less than 1000 s$^{-2}$ for weak confinement up to 15000 s$^{-2}$ for strong confinement. Previous experiments by Hebner and Riley [30] measured the wake attraction to be ≈ 200 fN on a lower particle at a horizontal displacement of 0.3 mm for operating parameters of 1.8 W and 60 mTorr, equivalent to weak confinement in our case (inset in Fig. 2). Employing a harmonic force approximation, this corresponds to an interaction strength of ≈ 1300 s$^{-2}$ *on* the $z$ axis, in reasonable agreement with the current measurement under similar conditions. A second result of interest is the high value of $D_{12x} ≈ 15000$ s$^{-2}$ found for high confinement strength. This is approximately 15 times the external horizontal confinement strength on particle 2 ($\omega_{2x}^2 ≈ 1000$ s$^{-2}$, as measured from the SMS method using Eq. (8)). This provides an explanation for why a vertical pair structure remains stable even though the external vertical confinement $\omega_{2z}^2$ (≈ 18000 s$^{-2}$) is much stronger than the horizontal confinement $\omega_{2x}^2$.

In both theory and simulation, the attractive horizontal wake force has been studied primarily for a downstream particle displaced from the $z$ axis, and in most studies, only the qualitative characteristic of the wake profile is discussed. On the $z$ axis, this force is zero due to symmetry, and the strength (force derivative) is sensitive to parameters such as the ion temperature $T_i$ (due to Landau damping). In fact, with the cold ion approximation, the wake potential diverges as it approaches the $z$ axis. Despite these complexities, comparison of our results with simulations of the ion wake using the COPTIC(SCEPTIC) PIC code, which takes into account the finite ion temperature $T_i$, still shows quantitative agreement: for experimental parameters similar to those used here, the magnitude of the horizontal repulsive force acting on the upstream particle is about 5 times less than that of the attractive force acting on the lower particle [24]. The results for $D_{12x}$ and $D_{21x}$ in Region I, as shown in Fig. 2, show excellent agreement with this theoretical prediction.

## IV. DISCUSSION

The asymmetric Yukawa-type upstream potential with different screening lengths $\lambda_z$ and $\lambda_x$ found in Section III (a) has been predicted previously from numerical calculations of the ion wakefield [25], in which the potential around a particle was calculated using linear response theory. In this case, the shifted Maxwellian ion velocities are described by a longitudinal dielectric function which includes Bhatnagar-Gross-Krook-type ion-neutral collisions [25]. While $\lambda_z$ and $\lambda_x$ are approximately equal to one another (and to the electron Debye length $\lambda_e$) for supersonic ion flow $M \geq 1$, they are shown to be different with $\lambda_z > \lambda_x$ for subsonic ion flow $M < 1$, approaching a limiting value equal to the ion Debye length $\lambda_i$ at $M \approx 0$ [25]. Here M is the Mach number defined by $M \equiv u_i/u_s$, where $u_i$ is the ion flow speed and $u_s$ is the ion sound speed. Thus, the experimental values determined here for $\beta$ in Region II, where $\beta > 1$ ($\lambda_z > \lambda_x$), are in agreement with this prediction.

Note that the calculation of $\beta$ only relies on the assumption of an effective Yukawa potential with different $\lambda_z$ and $\lambda_x$, but the detailed relation between M and $\beta$ is sensitive to the specific theoretical model used for the ion wake potential. Therefore, although the results reported in [25] may be used to roughly estimate the Mach number, for example, a value of $\beta \approx 1.7$ at $\omega_{-z}^2 \approx 2000$ s$^{-1}$ corresponds to a Mach number $M \approx 0.6$, a more sophisticated theory or a simulation more closely mimicking experimental conditions (for example, including the sheath inhomogeneity) will be needed for accurate measurements of the Mach number,

The transition of ion flow within the glass box from a supersonic to subsonic state as the rf power decreases is an interesting phenomenon, and is related to the modification of the plasma profile by the box used for horizontal confinement. Research using freely falling particles to map the electric field in the box [47] has shown that with decreasing rf power, the plasma profile transitions from a regular sheath, defined by an obvious sheath edge at which the Bohm criteria $M \approx 1$ is obeyed, to a "stretched" sheath exhibiting a more homogeneous profile. (Details of this research will be published in [48]). The results from current measurements and analysis clearly point to a subsonic ion flow corresponding to this "stretched" profile.

The phenomenon of a ratio of interaction strengths $|D_{12z}/D_{21z}|$ (and $D_w/D_D$) that is constant in Region I and increases in Region II, as found in Section III (b), can best be explained by considering the dependence of $\phi_w/\phi_D$ (equal to $D_w/D_D$) on M and $r/\lambda_e$. When $M \approx 1$ (Region I), it can be seen in Eqs. (17) and (18) that the ratio $\phi_w/\phi_D$ is a function of $r/\lambda_e$ only, remaining constant as long as the ratio $r/\lambda_e$ is constant. This can be seen to be roughly true in Fig. 4, where $\lambda_x$ and $\lambda_z$ ($\lambda_x \approx \lambda_z \approx \lambda_e$ when $M \approx 1$), within experimental error, exhibit a trend commensurate with that of $r$ in Region I. (A more strict proportionality between $r$ and $\lambda_e$ under varying rf power was reported earlier for vertical particle pairs in a regular plasma sheath without modification from a box [43].) The constant ratio between the interparticle spacing and the length scale of the potential can be conceptually understood as describing the situation when the particles are "pinned" in a given location within the potential profile generated by one another, even as their separation changes.

When M decreases, as occurs in Region II, $\phi_w$ is predicted to increase in magnitude and decrease in length scale by both the cold ion theory [15, 16] and simulations including ion-neutral collisions and a finite ion temperature (Landau damping) [25]. In this case, the length scale can be represented as $M\lambda_e$. At the same time, the magnitude of $\phi_D$, being a classical Yukawa potential, remains constant [15] while its length scale, represented by the vertical screening length $\lambda_z$, is also found to be approximately equal to $M\lambda_e$ [25]. Since the length scale $\lambda_z$ exhibits an increasing trend commensurate with that of $r$ in Region II (Fig. 4), the particles may again be considered to be "pinned" in the potential profile of one another. Therefore, the ratio $\phi_w/\phi_D$ (which equals $D_w/D_D$) depends on the magnitudes of $\phi_w$ and $\phi_D$ only. The increase in $D_w/D_D$ can then be explained by the fact that a decrease of M enhances the magnitude of $\phi_w$, but not that of $\phi_D$.

It is important to point out the difference between the explanation for the increase in $D_w/D_D$ reported in this research and that reported in [27], where two particles of different material (MF and polymethylmethacrylate (PMMA)) receded from one another as the lower particle lost mass due to its higher etching rate, while the plasma parameters were held fixed. In the latter case, the increase in $D_w/D_D$ can be explained by the increase in particle spacing $r$ surpassing the screening length $\lambda_z$.

## V. SUMMARY AND OUTLOOK

In this research, a non-perturbative method that we refer to as the scanning mode spectra (SMS) method is introduced that allows measurement of the interaction strengths and *in situ* confinement for a vertically aligned dust particle pair in a complex plasma. This method uses only the particles' thermal motion and the SMS is obtained by projecting particle thermal velocity onto eigenvectors that are scanned over the eigenvector space. Both the frequency and eigenvectors for the primary modes of oscillation can then be measured directly from the maxima in the SMS. These quantities are then used to simultaneously calculate the interaction strengths and *in situ* confinement in both the vertical and horizontal directions as derived from a coupled oscillator model theory (Eqs. (1-8)).

The SMS method is validated by obtaining results that are in agreement with those obtained from other experiments using different methods [26, 30, 40]. Based on the SMS obtained at a specific power (Fig. 1), the interaction between particles is found to be mutually repulsive but non-reciprocal in the vertical direction. In the horizontal direction, the force exerted by the upstream particle on the downstream particle is found to be attractive, while the force acting on the upstream particle from the downstream particle is repulsive. The lower particle is found to be heated and its charge reduced due to the upper particle's wake. The values of the interaction strengths in the vertical and horizontal direction, as well as the degrees of decharging and heating, agree quantitatively with previous experimental results [26, 30, 40]. The SMS method also improves upon and provides new insight into some of these results. For example, it allows the wake induced heating to be quantitatively represented by the ratios of the amplitude of the motion for the upper and lower particles $|\sigma_{\pm z}|$ and $|\sigma_{\pm x}|$, which are greater than one, and allows measurement of the horizontal wake attraction strength *on* the z axis.

The SMS method was then applied to measure the interaction of dust particle pairs at varying confinement strength $\omega_{-z}^2$ through variation of the rf power. By investigating the ratios among the interaction strengths in the horizontal and vertical directions, as well as in the upstream and downstream directions, theoretical and numerical predictions for the ion wake potential [15, 16, 20, 24, 25] are verified. The difference in the behavior of the ratio $|D_{21z}/D_{21x}|$ between the upstream strengths in the horizontal and vertical directions identifies two regions of the confinement strength $\omega_{-z}^2$ (and rf power): Region I (the high power region) and II (the low power region), with a boundary at $\omega_{-z}^2 \approx 6000$ s$^{-1}$ (power ≈ 3.6W). The upstream potential of a particle was found to be symmetric (with approximately equal screening lengths $\lambda_z$ and $\lambda_x$ in the vertical and horizontal directions) in Region I but asymmetric ($\beta = \lambda_z/\lambda_x > 1$) in Region II, with $\beta$ increasing for weaker vertical confinement (lower power) (Figs. 3 and 4). This finding verified predictions from previous simulations [25] and implies a transition of ion flow within the glass box from a supersonic to subsonic regime as the rf power decreases. Similarly, the ratio of the upstream to downstream interaction strengths in the vertical direction $|D_{12z}/D_{21z}|$ is found to be constant in Region I but increases as the power decreases in Region II (Fig. 5), which agrees with the theoretical prediction [15, 16, 25] that a decrease of Mach number enhances the magnitude of the ion wake potential. Finally, the ratio of downstream and upstream strengths in the horizontal direction $|D_{12x}/D_{21x}|$ is found to be a value ≈5 at high power, again verifying results predicted by numerical simulations [24].

Since the ion flow speed (Mach number) affects many fundamental properties and dynamic behaviors of complex plasmas, such as charging, dust structure and wave properties, the ability to tune the Mach number potentially allows the glass box (or any other designed confinement) to become a very useful tool for complex plasma research. Furthermore, due to the non-perturbative nature of this method, its application is not limited to complex plasmas, but can be applied to any system for which the thermal motion of the particles can be tracked, such as colloidal suspensions and granular materials. Extension of the method to more complex structures with additional constituent particles should also be possible upon application of more sophisticated spectra analysis that can deal with multiple dimensions.

## ACKNOWLEDGEMENTS

This work was supported in part by the National Science Foundation (NSF) under Grant No. PHY 1414523, 1707215, 1740203 and NASA contract 1571701.